\documentclass[aps,prb,twocolumn]{revtex4-1}

\usepackage{amsmath}
\usepackage{graphicx}
\usepackage{verbatim}
\usepackage{color}
\usepackage{hyperref}
\usepackage{array}

\newcommand{\head}[1]{\textrm{#1}}
\newcolumntype{C}[1]{>{\centering\arraybackslash}p{#1}}
\newcolumntype{L}[1]{>{\raggedright\arraybackslash}p{#1}}
\newcolumntype{R}[1]{>{\raggedleft\arraybackslash}p{#1}}

\definecolor{green}{RGB}{0,175,0}

\newcommand{\exciting}{{\usefont{T1}{lmtt}{b}{n}exciting}}

\begin{document}
\title{Energy-level alignment at organic/inorganic interfaces from first principles: \\ Example of poly(\emph{para}-phenylene) / rock-salt ZnO(100)}
\author{Dmitrii Nabok}
\author{Benjamin H\"{o}ffling}
\author{Claudia Draxl}
\affiliation{Institut f\"{u}r Physik and IRIS Adlershof, Humboldt-Universit\"{a}t zu Berlin, 12489 Berlin, Germany}

\begin{abstract}
By means of full-potential all-electron density-functional theory and many-body perturbation theory, we compute the band alignment at a prototypical hybrid inorganic/organic interface. The electronic properties of a model system built of poly(\emph{para}-phenylene) and \emph{rs}-ZnO are studied in two different geometries, employing several approaches of increasing sophistication. To this extent, we explore models for predicting the level alignment, which are based on the knowledge of the electronic structure of the individual constituents and are commonly used for semiconductor interfaces. For their evaluation in the context of hybrid materials, we perform an \textit{ab-initio} study of the entire system, including a quasiparticle description of the electronic stucture within the $G_0W_0$ approximation. Based on this, we quantify the impact of structure, charge redistribution, orbital hybridization, and molecular polarization on the band offsets and the alignment type. We highlight not only known limitations of predicting the level alignment at a hybrid inorganic/organic interface by simple models, but also demonstrate how structural details of the interface components impact the results.
\end{abstract}
\maketitle

\section{Introduction}
\label{sec:intro}
The electronic properties at the interface between active materials play a decisive role in determining the efficiency and functionality of electronic and opto-electronic devices. Understanding and tuning such systems is thus of outermost importance, which requires insight from microscopic simulations. The prediction of electronic band offsets from first principles remains, however, one of the great challenges of electronic-structure theory. This is in particular so when the two components forming the interface exhibit different nature, like hybrid inorganic/organic systems (HIOS).\cite{Draxl2014} Such hybrid interfaces exhibit a plethora of complex and even new phenomena, taking place between the substrate and the adsorbates.\cite{Ishii1999,Braun2009,Otero2017,Hofmann2019}
The fundamental physical processes that determine the energetics of HIOS, such as the Pauli-pushback effect, charge redistribution, and charge transfer, make it challenging to model such interfaces with simple mesoscopic and microscopic alignment methods. Hybrid materials should thus ideally be studied from first principles. However, despite enormous progress in the field, reliable \textit{ab-initio} calculations are often still out of reach.
On the one hand, one faces the problem that most popular density functionals are not applicable in such cases, and the most sophisticated electronic-structure theories need to be employed, such as the $GW$ approach based on a reasonable starting point [beyond semi-local functionals of density-functional theory (DFT)] or optimally tuned range-separated hybrid functionals\cite{Kronik2018}. On the other hand, the complexity of interfaces and heterostructures is likely to hamper the usage of highly accurate approaches. As a full many-body treatment of complex interface structures is often computationally unfeasible, many studies have to make use of approximations.

While molecules on metals have been studied more frequently, results for interfaces formed between inorganic and organic semiconductors are still scarce (for a recent review, see Ref.~\onlinecite{Hofmann2019} and references therein). 
For the latter, it is interesting to know to which extent one can apply level alignment schemes that have been widely employed for inorganic heterostructures; the most prominent examples being the Shockley-Anderson model,\cite{Anderson1962,Bechstedt1988} the branch-point alignment,\cite{Moench2004,Tersoff1984,Flores1979}, or the microscopic alignment via the electrostatic potential.\cite{Walle1987,Lambrecht1988,Lambrecht1990} 

An interface formed by poly-(para-phenylene) (PPP) and the rock-salt \emph{rs}-ZnO(100) surface is chosen in this work as a prototypical hybrid inorganic/organic system. Due to the small lattice mismatch and thus computationally affordable simulation cell size, this system is ideally suited for a theoretical in-depth analysis. We treat the interface in two geometries: (i) as an isolated PPP monolayer on a ZnO(100) surface, and (ii) as a heterostructure where a PPP monolayer on a ZnO(100) thin-film slab are periodically repeated along the surface normal. Our goal is to quantify the impact of finite-size effects, charge redistribution, molecular polarization, and orbital hybridization on the band offsets. For doing so, we start our exploration from a simple model. In the Shockley-Anderson approach, i.e., vacuum level alignment, one merely relies on the electronic properties of the constituents. In a more elaborate scheme, the band offsets are computed by aligning the microscopic electrostatic potentials of the individual systems with that of the entire interface. This treatment essentially still relies on the knowledge of the electronic properties of the individual materials. In this respect, we examine how sensitive the results are to the geometry of the individual systems. To this extent, we consider two configurations, i.e., the band edges of an isolated PPP chain and those of a PPP monolayer (termed PPP-ML) on the organic side, and another two systems, i.e. a ZnO bulk crystal and a 5-layer ZnO slab on the inorganic side. Eventually, the model results are compared with the corresponding data obtained from the band structure  of the entire interface. For this purpose we employ the $G_0W_0$ approximation of many-body perturbation theory.\cite{Hedin1965}

\section{Methods}

\subsection{Level-alignment}
\label{sec:alignment}
In the Shockley-Anderson model, the electronic levels are aligned with respect to the vacuum level, $E_{vac}$.
The band offsets are predicted as the differences between the electronic surface barriers, i.~e. ionization energies, $I_i=E_{\rm vac}-E_{\rm v,i}$, and electron affinities, $A_i=E_{\rm vac}-E_{\rm c,i}$, where $E_{\rm v,i}$ and $E_{\rm c,i}$ denote the respective valence band maximum (VBM) and conduction band minimum (CBm) of the constituent material, labeled $i$.
The electronic band offsets between two materials are then defined as
\begin{align}
\Delta E_{\rm v} & = I_2-I_1 = E_{\rm v,1}-E_{\rm v,2} \\
\Delta E_{\rm c} & = A_1-A_2 = E_{\rm c,2}-E_{\rm c,1}\mbox{.}
\label{eq:SA}
\end{align}
The signs are chosen such that a positive value means a potential barrier for the respective charge carrier (hole or electron).
Since this method derives interface properties from the properties of pristine materials, it neglects all interface contributions, such as interface dipoles, molecular polarization caused by substrate (or mutual) screening, or changes in the fundamental band gap due to orbital hybridization.
Nonetheless this method is still widely used in the search for promising interface materials for technological applications.
For probing its applicability, below we use this model in several variants, by obtaining its ingredients -- vacuum level, VBM and CBm -- from methods of increasing sophistication.

A more elaborate way of aligning electronic bands of the individual components at an interface / heterojunction beyond the Shockley-Anderson alignment, is to use the electrostatic potential as a reference level. This microscopic alignment technique accounts for all electrostatic effects at the interface, i.~e. charge rearrangements upon interface formation and interface dipoles. We will make use of this approach in a next step. However, changes in the electronic structure of the constituents through hybridization between their orbitals as well as non-local polarization effects of the molecular adsorbate by the underlying inorganic substrate are not considered in this procedure. To address these issues, a many-body treatment of the combined material is required. Therefore, we finally investigate the electronic structure of the heterostructure by a full many-body treatment, which also includes the influence of orbital hybridization and molecular polarization. In all the steps, we determine the electrostatic potential, $V_c$, and thus the vacuum level, $E_{vac}$ from self-consistent semi-local DFT calculations, while the electronic bands, and thus $E_v$ and $E_c$, are obtained from $G_0W_0$ calculations.

\subsection{Structural interface model}
\label{sec:StructureModel}
The alignment of the band structures via the electrostatic potential requires a structural interface model.
In the absence of experimental data on the atomic structure of the interface, the lattice-coincidence method is an appropriate tool to find stable geometries for heterostructures containing two periodic materials.\cite{Bechstedt1988,Koda2016} Good lattice coincidence means a minimum of dangling bonds but also small supercell size. The goal is to find a geometry with low strain and a supercell size that is computationally affordable. In the case of PPP and \emph{rs}-ZnO, we find the lattice constant of the polymer (8.159~bohr) to fit very well to the cubic axis of the oxide (8.198~bohr), resulting in lattice strain of less than 0.5\%.

\begin{figure}[tbp]
\includegraphics[width=1.0\columnwidth]{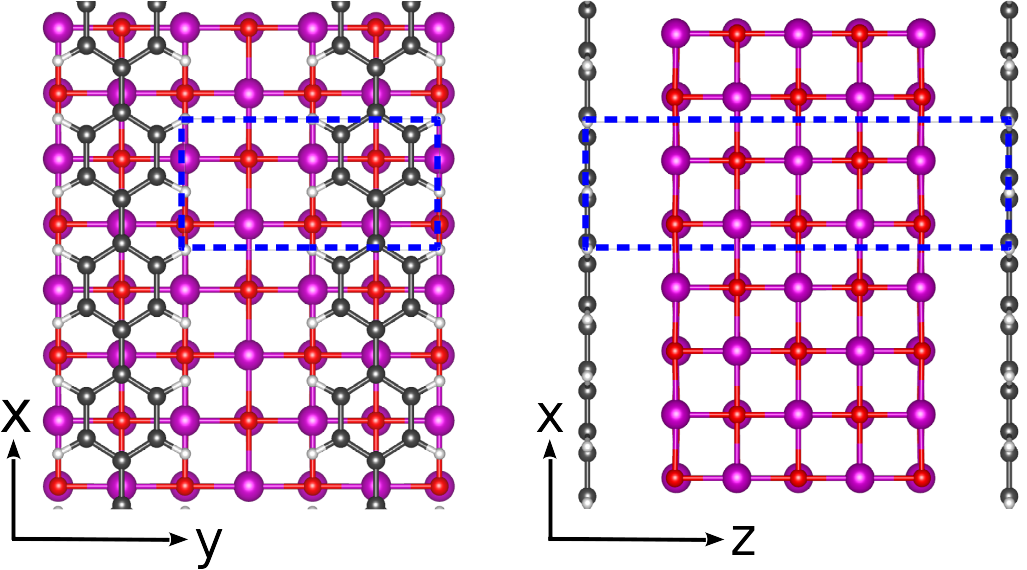}
\caption{\label{fig:interface-structure}Top and side view of a heterostructure consisting of a PPP monolayer and a thin ZnO film. Violet, red, gray, and white spheres indicate Zn, O, C, and H atoms, respectively. The unit cell is indicated by a blue dashed line.}
\end{figure}

For comparison, we  consider two types of interfaces: (i) the one that is formed between a PPP monolayer and a single ZnO(100) surface; and (ii) a heterostructure, where PPP is ``sandwiched'' between ZnO slabs.
In both cases, the unit cell, containing 50 atoms, consists of one ring of the flat-lying polymer, periodically repeated along the x-direction, and five layers of ZnO. The equilibrium adsorption geometry is considered to be the same for both interfaces, as obtained from crystal structure optimization performed for the heterostructure and presented in Fig.~\ref{fig:interface-structure}. Thus, the only difference between (i) and (ii) is the packing along the z-direction. The optimal distance between the two entities is about 5.858~bohr, in good agreement with characteristic adsorption distances upon physisorption.
The PPP chains are located at a distance of 8.676~\AA{} to each other, such that the minimum separation between neighboring atoms belonging to this PPP-ML is about 8.126~bohr. A similar surface adsorption geometry of PPP at the Cu(110) surface has been recently characterized by means of combined experimental and theoretical techniques.\cite{Vasseur2016}
The properties of the individual constituents have been computed employing a supercell geometry by introducing a sufficiently large vacuum layer, such to break the interaction between the periodic replicas. Thereby, the internal structure of the isolated PPP chain as well as the PPP-ML are kept the same as in the interface geometry. The Zn(100) surface unit cell is obtained based on the bulk structure. For all structures the vacuum separation is chosen based on convergence tests and amounts to at least as 16~bohr.
%

\subsection{Computational details}
\label{sec:ComputationalDetails}
\setlength{\tabcolsep}{1pt}
\begin{table}[htbp]
\caption{\label{tab:ComputationalDetails} Computational parameters employed in the $G_0W_0$ calculations.}
\begin{tabular}{ L{2.5cm} | C{1.5cm} C{1.5cm} C{1.5cm} R{1.5cm} }
\hline\hline
System       & \head{Dimension}  & \head{Vacuum (bohr)} & \head{k-grid} & \head{\# bands} \\
\hline
PPP         & 1D        & 20            & $56\times1\times1$  & 200  \\
PPP-ML      & 2D        & 24            & $8\times4\times1$   & 1000 \\
ZnO         & 3D        & -             & $6\times6\times1$   & 200  \\
ZnO(100)    & 3D        & 40            & $12\times12\times1$ & 1000 \\
PPP@ZnO (h) & 3D        & -             & $8\times4\times1$   & 1000 \\
PPP@ZnO (s) & 3D        & 30            & $8\times4\times1$   & 1000 \\
\hline\hline
\end{tabular}
\end{table}

All calculations are performed using the all-electron full-potential code \exciting.\cite{exciting,Nabok2016} The ground-state properties are treated within the framework of density-functional theory, employing the linearized augmented planewaves plus local orbitals (LAPW+lo) method. Exchange and correlation energies are approximated by the generalized-gradient approximation (GGA) in the PBE flavor.\cite{Perdew1996} The muffin-tin (MT) radii ($R_{\mathrm{MT}}$) for Zn, O, C, and H are 2.2, 1.6, 1.2, and 0.8 bohr, respectively. The LAPW basis set size is determined by the dimensionless parameter $R^{\mathrm{min}}_{\mathrm{MT}} G_{\mathrm{max}}$ which is chosen 7.0 for ZnO and 3.5 for PPP and PPP@ZnO. Here, $R^{\mathrm{min}}_{\mathrm{MT}}$ is the smallest MT radius of species in the respective compound. Both $R^{\mathrm{min}}_{\mathrm{MT}} G_{\mathrm{max}}$ values correspond to an effective plane-wave energy cutoff $G_{\mathrm{max}}$ of 4.375 bohr$^{-1}$.
For the groundstate properties, Brilloiun-zone samplings are carried out with grid sizes of $12\times1\times1$, $8\times2\times1$, and $8\times2\times1$ for the PPP chain, the ZnO surface, and the PPP@ZnO interface, respectively. Structural optimizations are performed such that all remaining atomic forces are below 0.3 mHa/bohr. To account for the predominate van der Waals (vdW) nature of the binding, we apply the DFT-D2 correction method.\cite{Grimme2006}

The electronic properties are calculated by the $G_0W_0$ approximation,\cite{Hedin1965, Hybertsen1986} using the Kohn-Sham (KS) orbitals as a starting point. We note that intensive convergence studies had to be conducted for each subsystem to achieve precise QP bandstructures. In particular, for obtaining the QP levels of the isolated PPP and the PPP ML, we apply a Coulomb-truncation technique to reach convergence with respect to the vacuum size.\cite{Ismail-Beigi2006} This is necessary due to the non-locality of the involved operators as otherwise the vacuum separation used in the groundstate calculations would be far from sufficient to obtain the QP electronic structure of 1D and 2D systems.\cite{Rozzi2006,Ismail-Beigi2006} The computational parameters used in the $G_0W_0$ calculations for all systems are summarized in Table~\ref{tab:ComputationalDetails}. Details of our $GW$ implementation are provided in Ref.~\onlinecite{Nabok2016}. In short, the dynamically screened Coulomb potential is computed in the random-phase approximation (RPA). For the frequency-convolution we employ the imaginary-frequency formalism where the dielectric function and the correlation self-energy are computed on an imaginary frequency grid.\cite{Jiang2013} To solve the quasiparticle equation, the correlation self-energy is analytically continued to real frequencies using the Pad\'e approximant. 16 imaginary frequencies turned out sufficient to produce stable results for states around the Fermi energy. Our computational setups assure convergence of the VBM and CBm energies within 0.05~eV for PPP, PPP-ML, bulk ZnO, and ZnO(100) slab. The error estimate for the interface is checked to be within 0.1~eV.

\section{Results and Discussion}
\label{sec:results}

\subsection{Electronic structure of PPP and PPP-ML}
\label{sec:ppp-gw}
\begin{figure}[hb]
\includegraphics[width=0.75\columnwidth]{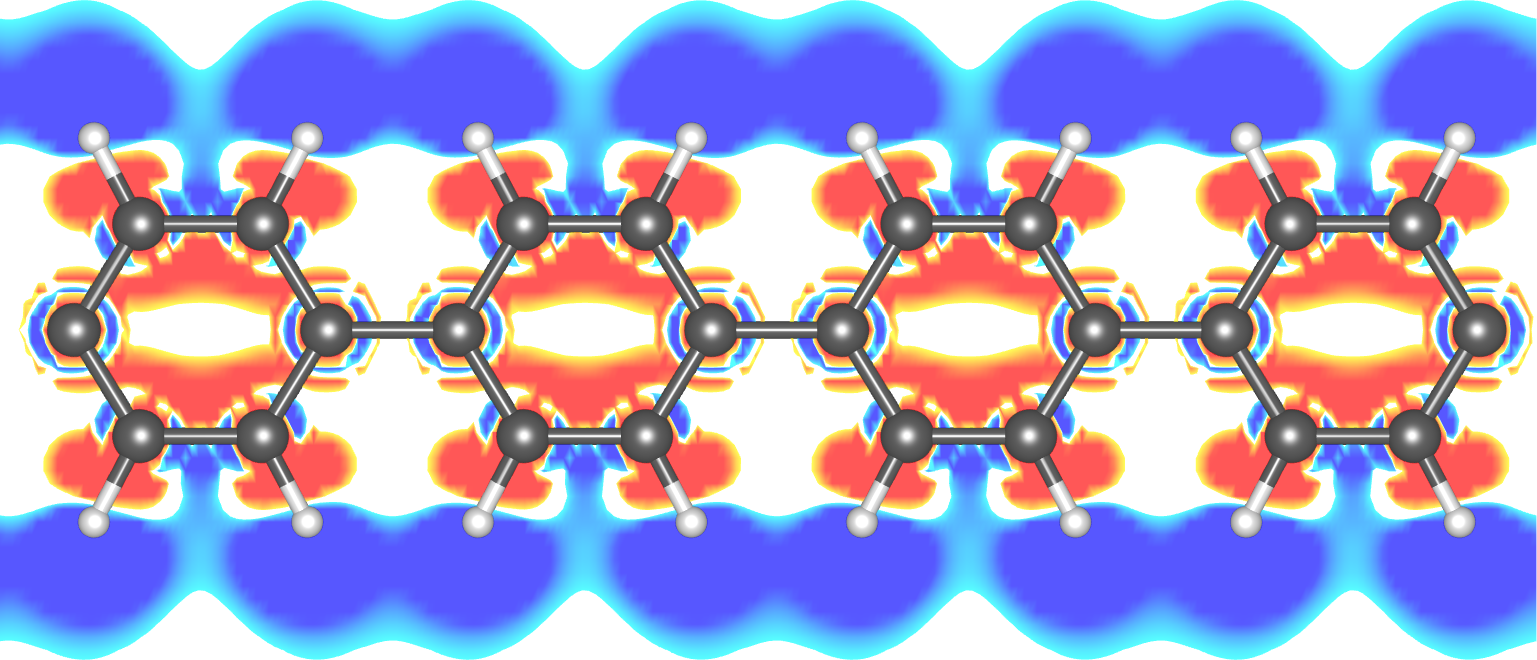}
\caption{\label{fig:ppp-rho-diff}Electron-density difference between a PPP monolayer and an isolated PPP chain (projected onto the polymer in the range $\pm 2\cdot10^{-5}$ e/bohr$^3$). Blue and red areas indicate charge accumulation and depletion, respectively. Clearly, upon ML formation, electron density is shifted towards the neighboring chain.}
\end{figure}

%
\setlength{\tabcolsep}{10pt}
\begin{table}[tbp]
\caption{\label{tab:ppp-IA} Electronic properties of PPP and PPP-ML.}
\begin{tabular}{ |l | c c | c c|}
\hline\hline
            & \multicolumn{2}{c|}{{\bf PPP}} & \multicolumn{2}{c|}{{\bf PPP-ML}} \\
            &  PBE    &  $G_0W_0$     &  PBE    &  $G_0W_0$        \\
\hline
$I$ (eV)   & -4.82   & -5.16         & -5.05   & -5.84            \\
$A$ (eV)   & -3.04   & -1.09         & -3.21   & -1.82            \\
$E_g$ (eV) &  1.77   &  4.07         &  1.84   &  4.02            \\
\hline\hline
\end{tabular}
\end{table}
The electron ionization energies ($I$), affinities ($A$), and fundamental band gaps of the 1D polymer and the 2D monolayer, as obtained from both DFT and $G_0W_0$, are presented in Table~\ref{tab:ppp-IA}. As a first observation, there is a pronounced change in these values going from the isolated chain to the monolayer, which amounts to $\approx$-0.2~eV for DFT and $\approx$-0.7~eV for $G_0W_0$. It is mainly an electrostatic effect that arises due to interaction of PPP chains and the thus enhanced polarizability of the $\pi$-electrons. As seen from the density-difference plot (Fig.~\ref{fig:ppp-rho-diff}), the intermolecular interaction causes a spread of the electron density. The sizable impact of monolayer formation stresses the sensitivity of any band-alignment model based on the quantities $I$ and $A$ to the details of system's geometry, as will be evident below.

Interesting to observe is also, how the band gap changes going from DFT to $G_0W_0$. The PBE value is by 70 meV larger in the ML than in the chain; however after applying the QP corrections the trend is changing to the opposite, and the value of the ML is getting 50 meV smaller than that of the isolated counterpart.
This rather small change is assigned to polarization effects, and overall, the values are very close to each other.

The experimental estimate for the optical band gap is in the range of 3.4-3.6 eV,\cite{Niko1996} based on optical-absorbance measurements and 4.0~eV from X-ray photoelectron spectroscopy.\cite{Riga1981} These values are in reasonable agreement with our computed ones, taking into account that there are several uncertainties that hamper direct comparison. On the one hand, the PBE starting point suggests that our $G_0W_0$ results are expected to be underestimated. On the other hand, we investigate geometries that serve the purpose of a model system, which concerns both the 1D as well as a selected ML. Moreover, we consider only the co-planar molecular geometry, although in equilibrium there is a torsion angle between nearest benzene rings which increases the band gap by roughly 0.4 eV on the DFT level.\cite{CAD1995} Overall, the computed band gaps are in very good agreement with DFT and $G_0W_0$ values reported in earlier studies.\cite{Artacho2004,Puschnig2012}

\subsection{Electronic structure of bulk \emph{rs}-ZnO and \emph{rs}-ZnO(100)}
\label{sec:zno-gw}
\begin{figure}[tbp]
\includegraphics[width=0.85\columnwidth]{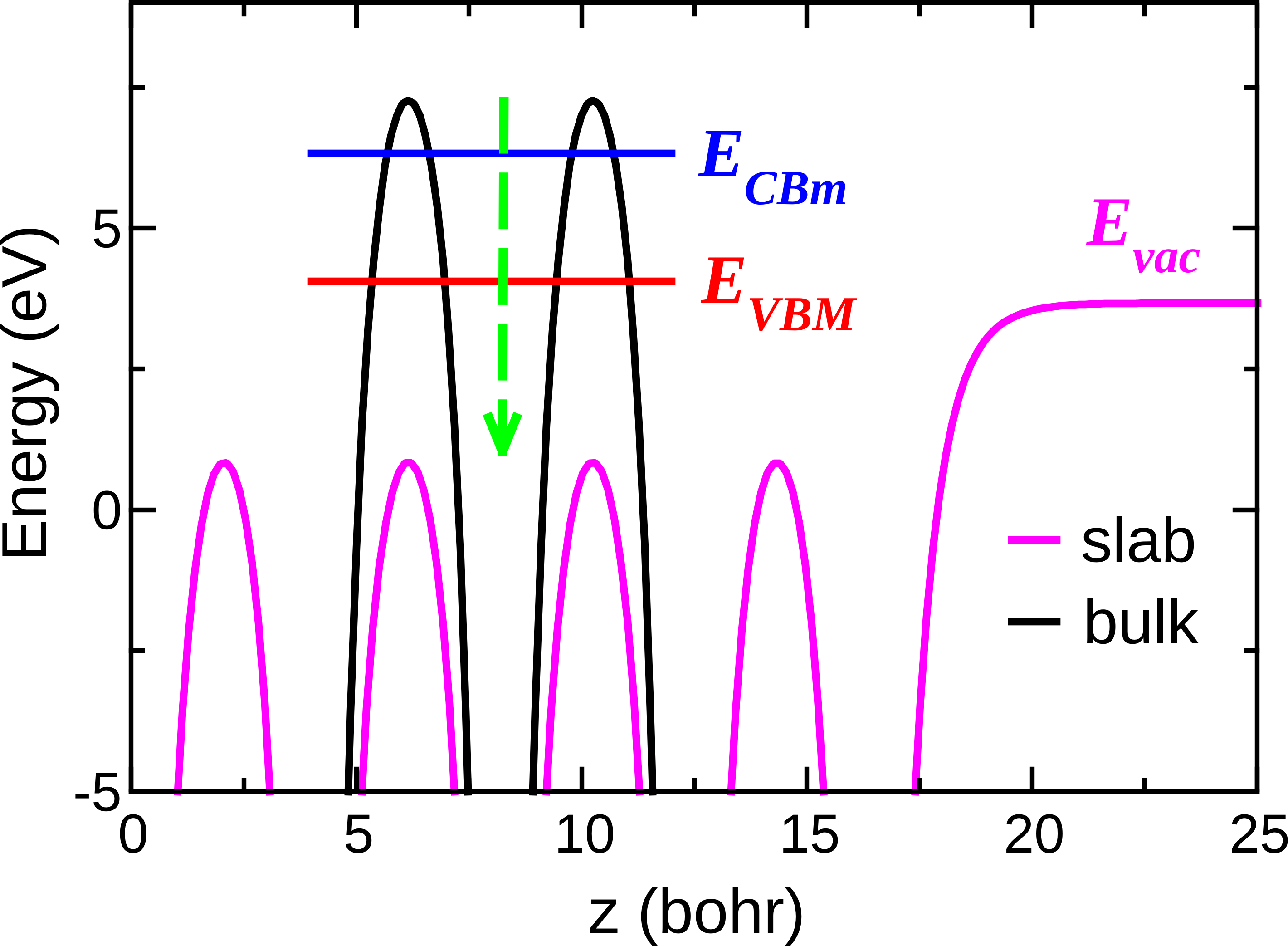}
\caption{\label{fig:zno-vc}
Procedure for estimating the electronic properties of the ZnO(100) surface based on ZnO-bulk data. The electrostatic potentials of the bulk structure (black) is shifted in energy such to fit the electrostatic potential of the ZnO(100) surface (magenta). This way one can use the surface ``vacuum'' energy and the correspondingly shifted values of bulk VBM and CBm to compute $I$ and $A$ of the surface.
}
\end{figure}

\setlength{\tabcolsep}{10pt}
\begin{table}[htb]
\caption{\label{tab:zno-IA} Electronic properties of \emph{rs}-ZnO(100) from two different scenarios. On the right, all values are computed for the surface, consisting of 5 MLs. On the left, $E_g$ is computed from $G_0W_0$ for the bulk, while $I$ and $A$ are obtained by aligning the middle layers of the electrostatic potential of the surface to that of the bulk.}
\begin{tabular}{ |l | c c | c c|}
\hline\hline
            & \multicolumn{2}{c|}{\bf bulk} & \multicolumn{2}{c|}{\bf slab} \\
            &  PBE    &  $G_0W_0$     &  PBE    &  $G_0W_0$         \\
\hline
$I$ (eV)    & -4.99   & -6.07         & -5.00   & -6.05             \\
$A$ (eV)    & -4.25   & -3.80         & -4.34   & -3.72             \\
$E_g$ (eV)  &  0.74   &  2.27         &  0.66   &  2.33             \\
\hline\hline
\end{tabular}
\end{table}
The electronic properties of the ZnO component are computed for bulk \emph{rs}-ZnO as well as a \emph{rs}-ZnO(100) surface using a 5-layer periodic slab model for the latter. To obtain insight to which extent the surface electronic structure can be understood by adopting quantities from the respective bulk material, we pursue two different approaches. In the first one, we estimate the QP energies of the slab using the $G_0W_0$ values obtained for bulk \emph{rs}-ZnO and aligning the electrostatic potentials of the bulk and the middle layers of the slab. This procedure is shown in Fig.~\ref{fig:zno-vc}.
In the second approach, we compute the QP values directly for the slab. The corresponding results are presented in Table~\ref{tab:zno-IA}.

Focusing on the bulk results first, one should note that the rock-salt polymorph is much less often studied compared to other ZnO phases. An experimental lattice constant of 4.28~\AA{}~\cite{Segura2003,Sans2005} has been reported, which is in between the theoretical LDA result (4.21~\AA) and GGA-PBE results of 4.334~\AA~\cite{Schleife2006} and 4.338 (this work). Overall, our Kohn-Sham band structure is almost identical to that of Ref.~\onlinecite{Schleife2006}. We obtain an indirect GGA-PBE band gap of 0.74~eV and a direct gap at $\Gamma$ of 1.96~eV. These results are, however, significantly different from LDA values obtained with optimized pseudopotentials by Dixit an coworkers,\cite{Dixit2010} who report a direct LDA gap of 2.79~eV. Considering this discrepancy as a major difference in the starting point, it is not surprising that also our G$_0$W$_0$ band gaps (2.26~eV (indirect) and 3.32~eV (direct) are significantly smaller than than the direct gap of 4.27~eV reported in Ref.~\onlinecite{Dixit2010}. In contrast, the estimated indirect quasiparticle gap of 2.6~eV, based on an analytical model~\cite{Bechstedt1988a} as provided in Ref.~\onlinecite{Schleife2006} is in reasonable agreement with our value.
Experimental values for the indirect and direct band gaps are reported to be 2.7~eV and 4.6~eV,\cite{Sans2005} respectively, based on the analysis of the pressure-dependent optical adsorption edges. In essence, our $G_0W_0$@PBE values underestimate the band gaps in the same way as it is known for other ZnO polymorphs. This pronounced starting-point dependence in oxides~\cite{Schleife2012} could be remedied by using the hybrid functionals. As extensive $GW$ calculations with such functionals are numerically not affordable for the hybrid materials, it is not in the focus of this work.

We proceed with discussing the electronic structure of the \textit{rs}-ZnO(100) surface. The results displayed in Table~\ref{tab:zno-IA}, obtained by either using the bulk electronic structure or the one computed for the actual surface geometry, are rather similar. Consequently, the surface band gap is also indirect with the same location of VBM. Like in the PPP case, we notice a different trend in the fundamental band gap between DFT and $G_0W_0$ when going from the more densely packed system to the surface. While the DFT value is 80~meV larger for the bulk crystal than for the slab, the corresponding $G_0W_0$ value is 60~meV smaller. At this point, we like to comment about how realistic our 5-layer slab model is for simulating the electronic structure of the ZnO(100) surface. By repeating the calculations using 7 layers, we find that both the DFT and $G_0W_0$ band gaps do not change drastically, with a difference of only 10~meV for DFT and 50~meV for $G_0W_0$.

\subsection{Shockley-Anderson lineup}
\label{sec:meso}
Having in hand the values for $I$ and $A$ of both components, one can readily apply the simplest Shockley-Anderson model to determine the level alignment and band discontinuities for our organic/inorganic interface.
In Table~\ref{tab:band-offsets}, we present results for 4 types of systems as obtained by combining the results for PPP and PPP-ML with ZnO(100)-bulk and ZnO(100)-slab, respectively, as described above.
\begin{table*}[btp]
\caption{\label{tab:band-offsets} Band gaps and electronic band offsets (in eV) corresponding to different models, computed with $G_0W_0$ and DFT (in parenthesis). ZnO(100)-bulk and ZnO(100)-slab denote the models where the QP bandstructure is obtained using bulk data or the 5-layer slab model. (h) and (s) indicate results for the heterostructure and surface system, respectively.}
\begin{tabular}{| l | c c l |}
\hline\hline
Model & $\Delta E_{\rm v}$ & $\Delta E_{\rm c}$ & Alignment \\
\hline
\multicolumn{4}{|c|}{{\bf Shockley-Anderson model}} \\
\hline
PPP/ZnO(100)-bulk    &  0.91 ( 0.18) & -2.71 (-1.21) & type II (II) \\
PPP/ZnO(100)-slab    &  0.89 ( 0.19) & -2.63 (-1.30) & type II (II) \\
PPP-ML/ZnO(100)-bulk &  0.23 (-0.05) & -1.97 (-1.04) & type II (I)  \\
PPP-ML/ZnO(100)-slab &  0.21 (-0.04) & -1.89 (-1.13) & type II (I)  \\
\hline
\multicolumn{4}{|c|}{{\bf Microscopic alignment model}} \\
\hline
PPP-ML/ZnO(100)-bulk (h) &  0.22 (-0.06) & -1.97 (-1.04) & type II (I)  \\
PPP-ML/ZnO(100)-slab (h) &  0.23 (-0.01) & -1.93 (-1.17) & type II (I)  \\
PPP-ML/ZnO(100)-bulk (s) &  0.02 (-0.26) & -1.77 (-0.84) & type II (I)  \\
PPP-ML/ZnO(100)-slab (s) &  0.03 (-0.23) & -1.71 (-0.95) & type II (I)  \\
\hline
\multicolumn{4}{|c|}{{\bf Interface band structure}} \\
\hline
PPP-ML/ZnO(100)-slab (h) & -0.21 (-0.42) & -0.09 (-0.21) & type I  (I)  \\
PPP-ML/ZnO(100)-slab (s) & -0.28 (-0.35) & -0.44 (-0.52) & type I  (I)  \\
\hline\hline
\end{tabular}
\end{table*}
The results reveal rather strong variations, depending on the system and method to obtain $I$ and $A$.
Comparing the DFT values, one can conclude that the major difference in the band discontinuities comes from the adsorbate, namely from the choice of adopting either an isolated chain or a monolayer. This results into an almost rigid shift of $\Delta E_{\rm v}$ ($\Delta E_{\rm c}$) by about -0.2~eV (+0.2~eV) and, most important, even a change of alignment type from II to I. Quasiparticle corrections, obtained by $G_0W_0$, result into an almost rigid shift of $\Delta E_v$ ($\Delta E_c$) by about -0.7~eV (+0.7~eV), however, do not change the type of the alignment between the two scenarios.

\subsection{Microscopic alignment model}
\label{sec:micro}
\begin{figure}[tbp]
\includegraphics[width=0.95\columnwidth]{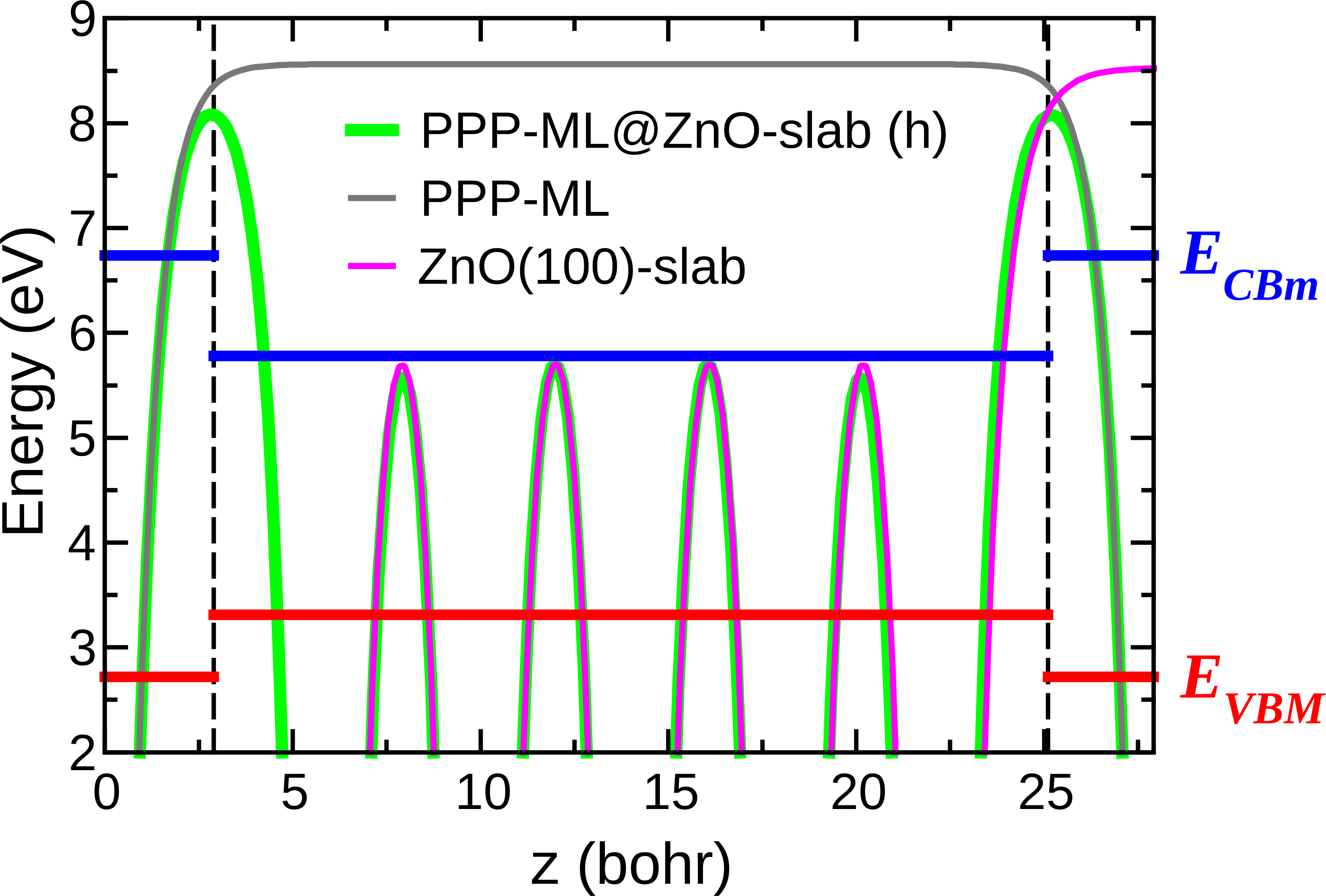}
\caption{\label{fig:micro-vc} Microscopic alignment model for computing the band discontinuities at the PPP-ML@ZnO(100)-slab (h) interface. The electrostatic potentials of the isolated PPP-ML (grey) and ZnO(100) surface (magenta) are aligned such to fit the electrostatic potential to that of the full interface (green). This way, energy shifts are obtained that are applied to $I$ and $A$ of the isolated subsystems for computing the band offsets.}
\end{figure}

\begin{figure*}[tbph]
\centering
\includegraphics[width=0.4\textwidth]{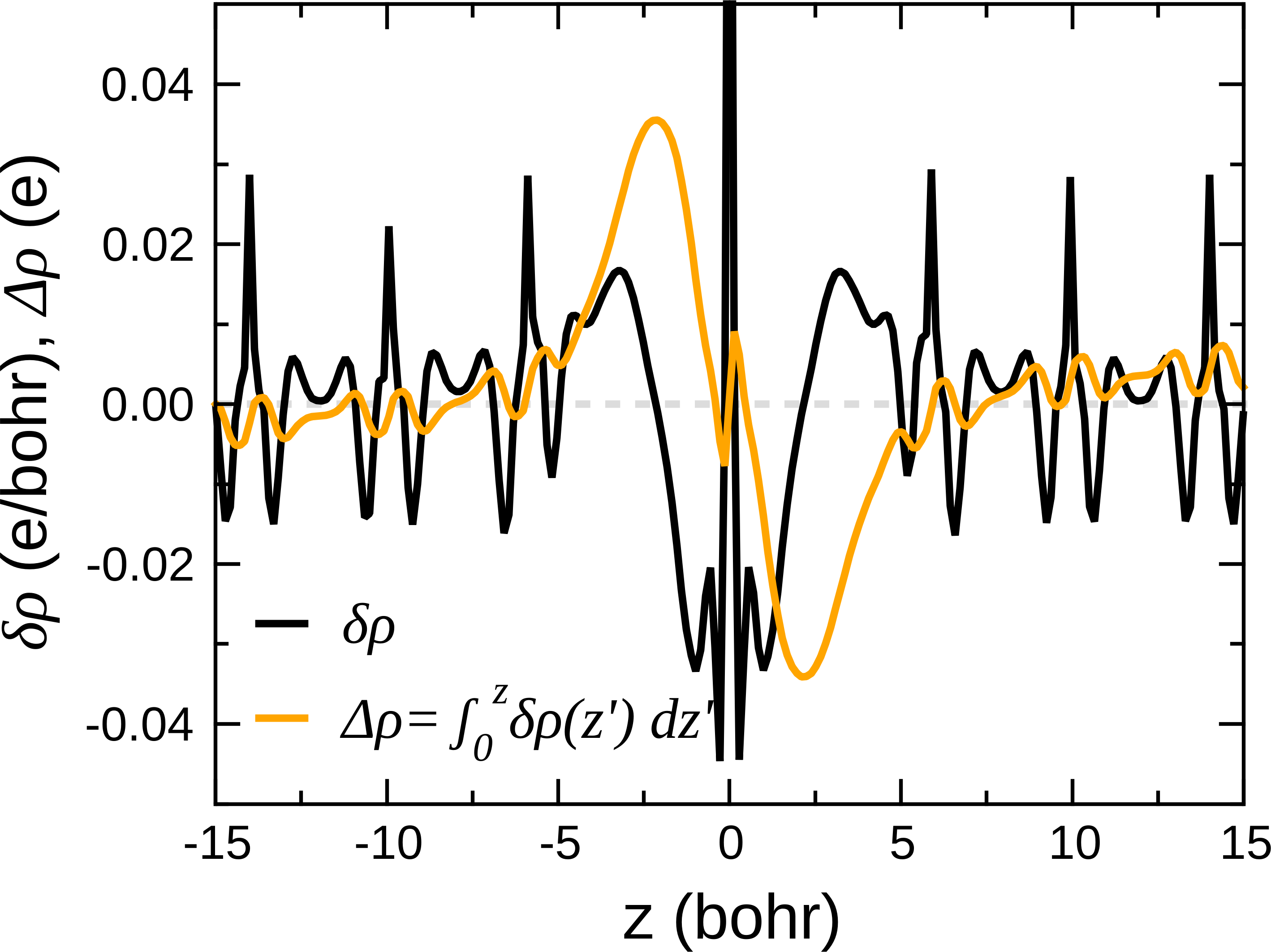}
\qquad
\includegraphics[width=0.4\textwidth]{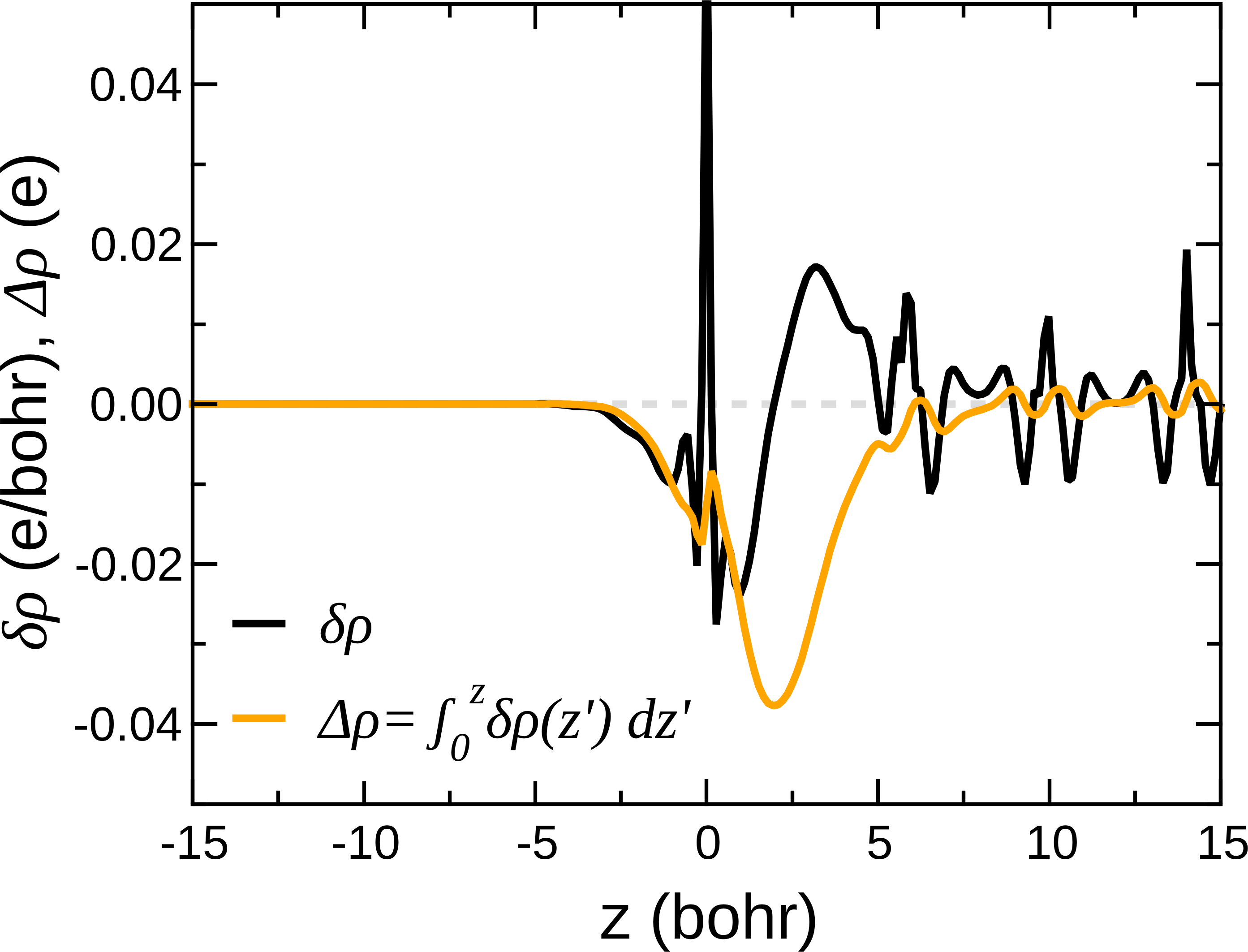}
\caption{\label{fig:micro-rho-vc-diff}Plane-averaged electron-density difference between the total density of the hybrid system and those of its constituent systems, for the heterostructure (left) and the surface geometry (right).}
\end{figure*}

The main limitation of the Shockley-Anderson model is the complete neglect of all effects that take place upon interface formation. To release this constraint requires a microscopic structural model of the interface. Therefore, we now proceed with another technique to compute the level alignment which is widely used for describing properties of inorganic semiconductor heterostructures. The microscopic alignment model~\cite{Walle1987} is designed to capture the electrostatic effects arising at the interface and predict the effective energy-level alignment based solely on the accurate electronic structure of the constituents. To this extent, we perform DFT calculations for both interface geometries described in Sec.~\ref{sec:StructureModel} to obtain the electrostatic potential of the relaxed combined systems, that are then aligned with the electrostatic potentials of its isolated constituents as shown in Fig.~\ref{fig:micro-vc}.
The resulting band offsets are presented in Table \ref{tab:band-offsets}. Depending on the interface geometry, we observe two different predictions. The values for the heterostructure are found to be almost identical to those of the Shockley-Anderson model, however the corresponding values for the surface are rigidly shifted by around 0.2 eV. The reason for this difference can be understood based on an analysis of the electron-density difference at the interface and the corresponding effect on the electrostatic potential, as presented in Fig.~\ref{fig:micro-rho-vc-diff}.
As it follows from the figure, the interface formation leads to a charge redistribution such that charge accumulation and depletion is taking place
predominately in the area between the molecule and the ZnO surface. This leads to the formation of an interface dipole, with the corresponding electric field causing a drop of the electrostatic potential in the area between the two constituent materials. This, in turn, leads to a significant modification of the band discontinuities with respect to the Shockley-Anderson model. This scenario also holds for the surface geometry (right panel), where the observed 0.2 eV shift is consistent with the solution of the Poisson equation for the shown density-difference profile. Looking at the results for the heterostructure, it becomes clear that due to the ``sandwich''-like packing geometry, the electrostatic effect is effectively canceled out due to presence of the second symmetric surface that causes the formation of the same electric field but in the opposite direction.
Lastly, similar to the Shockley-Andersen case, the results obtained based either on bulk data or on the surface slab configuration are found to be very similar in the case of the microscopic alignment model.

\subsection{Interface Band Structure}
\label{sec:gw}
%
\begin{figure*}[tbp]
\includegraphics[width=0.4\textwidth]{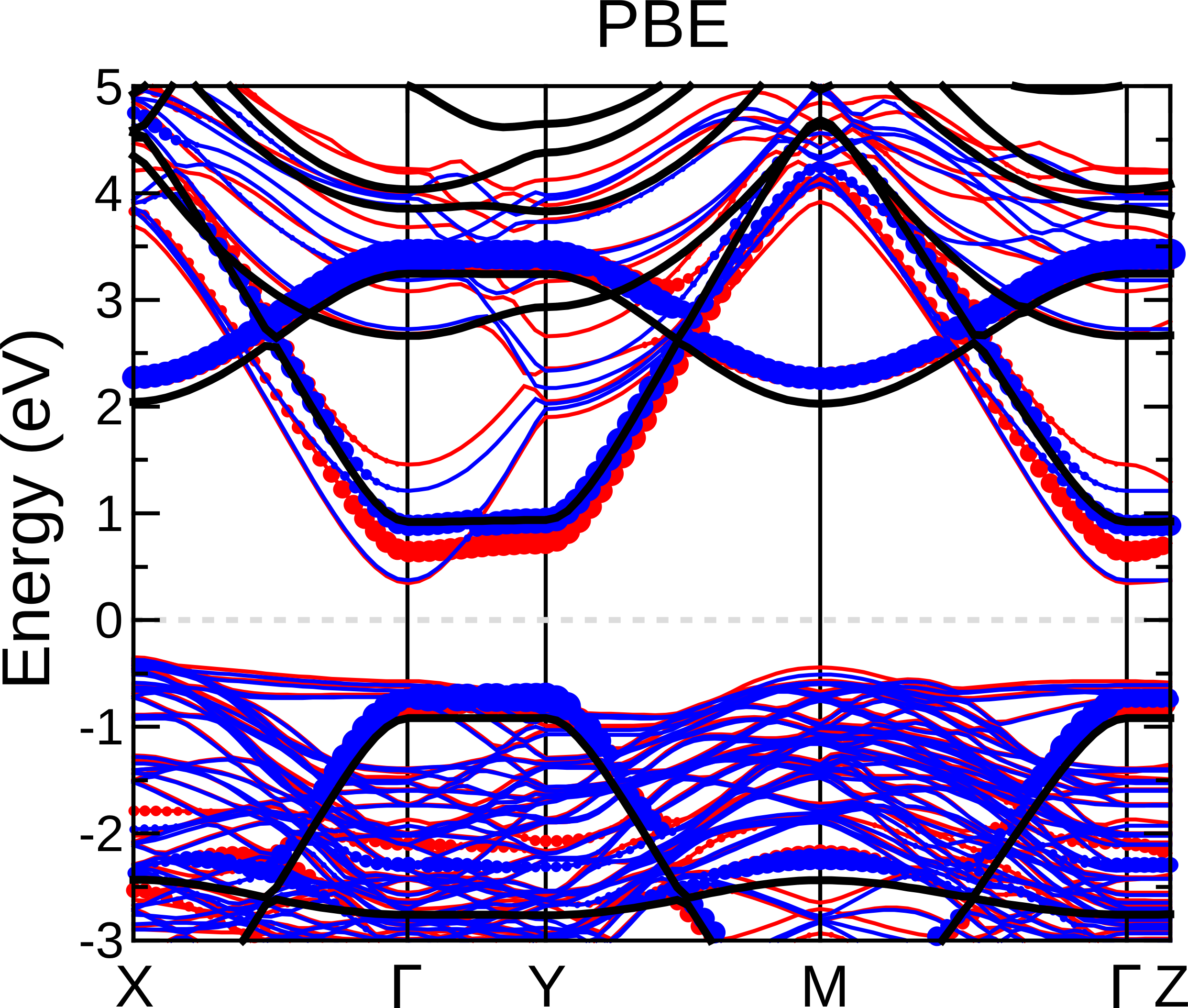}
\qquad
\includegraphics[width=0.4\textwidth]{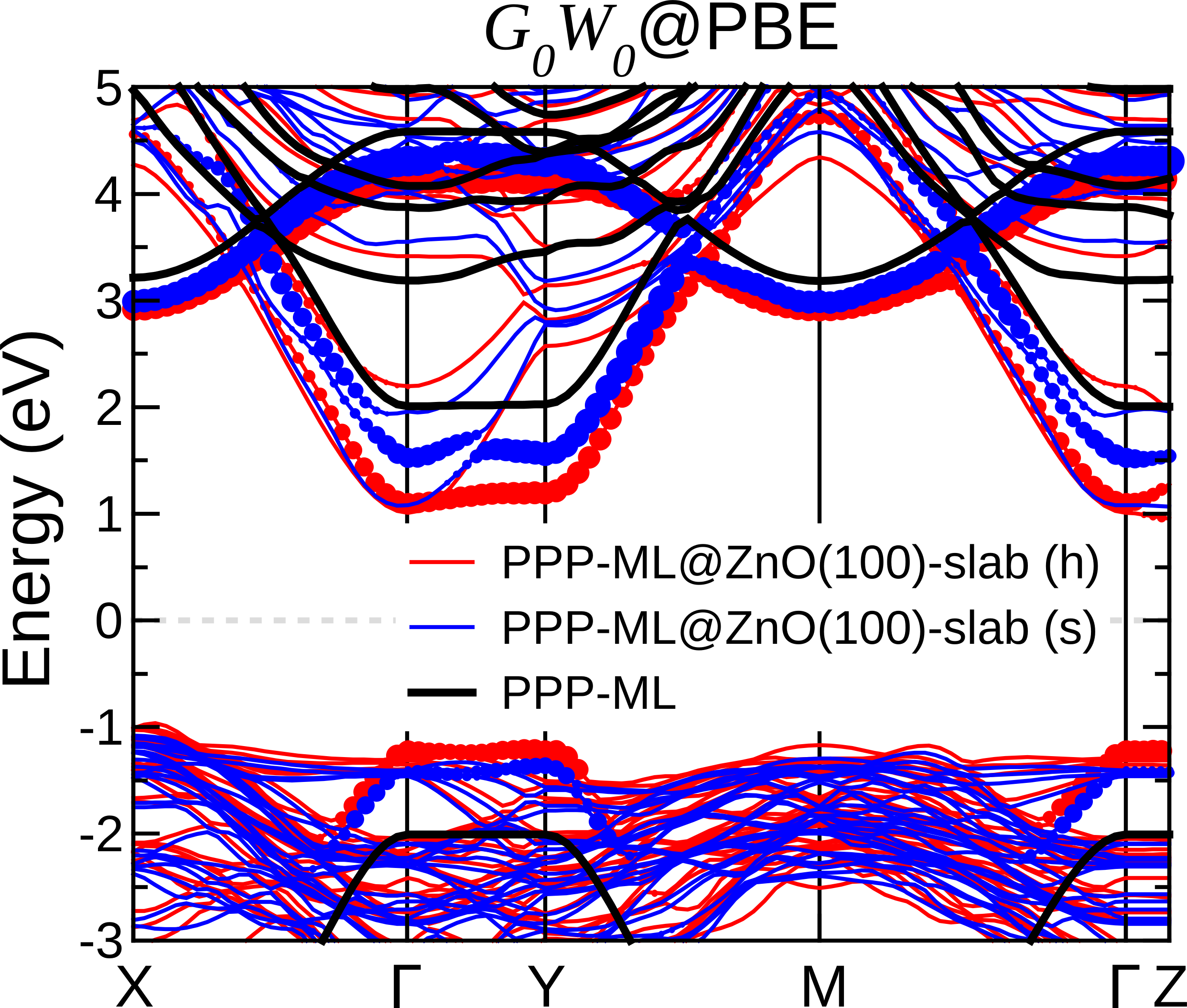}
\caption{\label{fig:bands}PBE (left panel) and $G_0W_0$@PBE band structures (right panel) for the PPP-ML@ZnO(100)-slab interface in the heterostructure and surface geometry. For comparison, the energy levels of the isolated PPP monolayer are also included. Circles denote the electronic states with a predominant molecular character. All band structures are aligned with respect to the Fermi energy, located in the middle of the corresponding band gap.}
\end{figure*}
As shown above, the microscopic treatment of the system has revealed important modifications in the electron density distribution upon interface formation. In this model, we still kept a major assumption, namely that the electronic properties of constituents are only weakly affected. In other words, we have neglected the effects of polarization and orbital hybridization on the band edges. In order to quantify these contributions, we proceed with an analysis of the QP bandstructure of the entire --interacting-- system.

\begin{figure}[thbp]
\includegraphics[width=0.95\columnwidth]{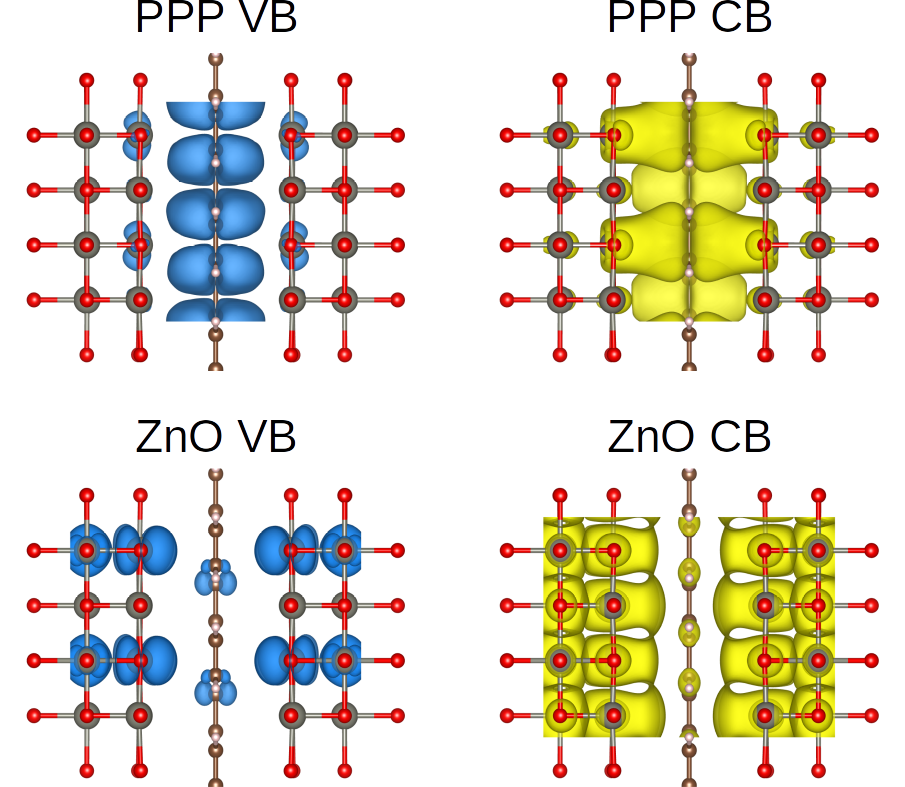}
\caption{\label{fig:wf3d}Kohn-Sham wavefunctions ($|\psi_{n\mathbf{k}}|$) at $\Gamma$ corresponding to the VBM and CBm of PPP-ML and ZnO(100). The isovalue is set to 0.0002 e/bohr$^3$.}
\end{figure}

Both the DFT and the $G_0W_0$ bandstructure diagrams are presented in Fig.~\ref{fig:bands} for the two discussed interface geometries. For comparison, the bands of the isolated PPP monolayer are also indicated (black lines). Analyzing the DFT bandstructure, first of all, one can clearly see the quasi one-dimensional dispersion of the polymer bands. Remarkably, these states are only weakly affected in the case of the surface geometry. However, the changes get more pronounced in the heterostructure geometry, where the hybridization of the PPP and ZnO states gives rise to a dispersion of the molecular bands along the $\Gamma$-Z and (though rather weak) $\Gamma$-Y directions. To further illustrate this point, the wavefunctions attributed to valence and conduction states of PPP-ML and ZnO(100) at the $\Gamma$ point are presented in Fig.~\ref{fig:wf3d}. One can see that the interaction with the substrate has a particularly strong effect on the shape of the conduction state of PPP (for simplicity called LUMO).

The effect of hybridization between the polymer $\pi$ and $\pi^{*}$ and the ZnO states can be quantified by comparing the HOMO-LUMO gaps \footnote{In the case of a combined system, the PPP HOMO and LUMO energies are determined by the analysis of VB and CB states with predominate molecular character.} of PPP in the different geometries. The DFT value reduces from 1.8 eV for the isolated case, to 1.6 eV in the surface geometry, and to 1.4 eV in the heterostructure. The further reduction by 0.2 eV in the latter case, is understood by the presence of the second ZnO surface in this geometry.

Applying QP corrections lead to even more pronounced changes in the electronic structure of the interface.
First of all, the QP corrections are strongly dependent on the nature of the electronic state.
Thus, due to their more localized nature, the PPP QP levels are renormalized more strongly than the ZnO states. The $G_0W_0$ band gap of the isolated PPP monolayer (4 eV) is significantly larger than the corresponding value in the surface geometry (2.9 eV) and the heterostructure (2.3 eV). Notably, the QP corrections also lead to a change in the order of states. For instance, the top ZnO and PPP-ML valence bands are swapped in the vicinity of the $\Gamma$ point. It is interesting to note that the ZnO band gap is less affected by the interface formation, also evident from the rather small changes in the electron density distribution inside of the ZnO slab (see Fig.~\ref{fig:micro-rho-vc-diff}). The renormalization of the ZnO band gap is only 0.03 eV in DFT but anyway 0.45 eV in $G_0W_0$. A qualitatively similar behavior has been recently observed in another hybrid system, consisting of pyridine molecules chemisorbed on the wurtzite ZnO($10\overline{1}0$) surface.\cite{Turkina2019}

The very strong renormalization of the PPP HOMO-LUMO gap at the interface is a combined action of two contributions. The first one is due to orbital hybridization which is already included at the DFT level. The second and dominant contribution stems from the polarization-induced renormalization of the molecular levels due to the presence of polarizable media in the vicinity of a molecule.\cite{Johnson1987,Neaton2006} Even though ZnO is a semiconductor, the contribution due to this effect is estimated to be 0.9 eV for the surface and 1.2 eV for the heterostructure. This is not too surprising since effects of a few tenths of an eV have even be observed due to mutual polarization of 2D materials.\cite{Fu2016} Our values for the polarization-induced reduction of the molecular band gap are close to those reported for similar systems.\cite{Garcia-Lastra2009,Puschnig2012}

We finally discuss our results for the band offsets. They are computed by identifying the character of valence and conduction states in the vicinity of the Fermi energy. The corresponding values are presented in Table~\ref{tab:band-offsets}. Very important, both DFT and $G_0W_0$ values for the entire interface differ significantly from the model predictions and even exhibit a different alignment type, changing from type II to type I. Remarkably, the values for $\Delta E_c$ are modified by more than 1.5 eV. It is also interesting to note that the DFT values differ only by 0.2 eV from the corresponding $G_0W_0$ energies. We regard this, however, a pure coincidence. Comparing the different adsorption geometries, the presence of the second ZnO surface reduces $\Delta E_c$ by a factor of 4, while $\Delta E_v$ is almost unchanged. This strong modification of $\Delta E_c$ can be attributed to the above discussed noticeable hybridization between the PPP and ZnO conduction states (Fig.~\ref{fig:wf3d}).


\section{Conclusions}

In summary, for a prototypical inorganic/organic hybrid material (PPP/rs-ZnO), we have probed various interface models, that potentially allow for predicting the interfacial electronic structure from the properties of its constituents, with the aim to quantify their errors. Overall, the considered models cannot provide a satisfactory description of the band offsets, even when taking into account the optimized geometries and the knowledge of the accurate electronic structures of the isolated subsystems. Most striking, the alignment type predicted by all models (type-II) is found to be different from the results of the microscopic treatment of the entire system (type-I). The relatively weak interaction between PPP and ZnO causes nevertheless a pronounced redistribution of the electron density at the interface, accompanied by a noticeable orbital hybridization between the LUMO of PPP and the ZnO conduction states. Accounting for this hybridization reduces the value of $\Delta E_c$ by more that 1~eV with respect to the model treatment. QP effects included via the $G_0W_0$ approximation have been shown to play an important role in the electronic structure of both, the isolated constituents and the hybrid material. On the one hand, it naturally provides improved values for  band gaps, ionization potentials, and electron affinities. On the other hand, the $G_0W_0$ approach has shown to be indispensable for quantifying the polarization-induced renormalization of the molecular electronic structure at the interface. Conversely, the properties of the PPP@ZnO interface are found to be very sensitive to the microscopic picture, requiring a treatment that is capable to capture the peculiar interplay between charge density redistribution, orbital hybridization, and polarization-induced band renormalization. For a reliable {\it ab initio} description of level alignment at organic/inorganic interfaces and heterostructures, a many-body approach is indispensable, or should at least be used to develop more appropriate models.

\section*{Acknowledgment}
Work supported by the Deutsche Forschungsgemeinschaft (DFG) - Projektnummer 182087777 - SFB 951.
All input and output files can be downloaded from the NOMAD Repository, http://dx.doi.org/10.17172/NOMAD/2019.07.12-1.

\bibliography{bibliography}

\end{document}